\documentstyle[12pt,aasms4] {article}
\newcommand{\beq}{\begin{equation}}
\newcommand{\eeq}{\end{equation}}
\newcommand{\beqn}{\begin{eqnarray}}
\newcommand{\eeqn}{\end{eqnarray}}

\newcommand{\pa}{\partial}

\newcommand{\non}{\nonumber}

\begin{document}
 
\title{\bf{Dynamical Effects from Asteroid Belts for Planetary Systems}}
\author{Ing-Guey Jiang$^{1}$  and Li-Chin Yeh$^{2}$}

\affil{{$^{1}$ Institute of Astronomy,}\\
{ National Central University, Chung-Li, 
Taiwan} \\  
\ \\
{$^{2}$ Department of Mathematics,}\\
{ National Hsinchu Teachers College, Hsin-Chu, Taiwan}}

\authoremail{jiang@astro.ncu.edu.tw}

\begin{abstract}
The orbital evolution and  stability of planetary systems 
with interaction from the belts is studied 
using the standard phase-plane analysis.
In addition to the fixed point which corresponds to the Keplerian orbit,
there are other fixed points around the inner and outer edges of the belt.
Our results show that for the planets,
the probability to move stably
around the inner edge is larger than the one to move around the outer edge.
It is also interesting that there is a limit cycle of semi-attractor for a 
particular case.
Applying our results to the Solar System, 
we find that our results could
provide a natural mechanism to  do the 
orbit rearrangement for the larger Kuiper Belt Objects and thus successfully 
explain the absence of these objects beyond 50 AU. 
\end{abstract}

\newpage

\section{Introduction}

Newton's Law of gravity and motion has been used as the most fundamental
law for the Physical Sciences since its success in explaining 
the motion of celestial
bodies in the Solar System.  Thus, the Newton's Law was in fact first proved 
in the astronomical context. It was then applied to other fields successfully. 


For example,
Clausen et al. [1998] studied periodic modes of motion of a few 
body system of magnetic holes, both experimentally and numerically.
Kaulakys et al. [1999] showed that a system of many bodies
moving with friction can experience a transition to chaotic behavior. 
Xia [1993] did some work on Arnold diffusion
in the elliptic restricted three-body problems. 

Many new discoveries of extrasolar planets have been made recently 
(Marcy et al. [1997]) and these
events really provide exciting and important opportunities to understand
the formation and evolution of planetary systems, including the Solar System.
According to the Extrasolar Planets Catalog 
(http://cfa-www.harvard.edu/planets/catalog.html) maintained by Jean Schneider,
117 planets have been detected till date (July 2003).
These planets have masses ranging from   
0.16 to 17 Jupiter mass ($M_J$) and semimajor  
axes ranging from 0.04 AU to 4.5 AU and a wide range
of eccentricities. Therefore, the dynamical properties of some of 
these planets are very 
different from the planets in the Solar System.

Nevertheless, there do exist similarities between the extrasolar and 
the Solar System planets. 
For example, there is a new discovery of a Jupiter-like orbit,
i.e. a Jupiter-mass planet on a circular long-period orbit 
(Marcy et al. [2002]). 
Moreover, some planetary systems are claimed to have discs of dust and
they are regarded to be young analogues of the Kuiper Belt in our Solar
System (Jayawardhana et al. [2000]).  

If these discs are massive enough, they should play important roles 
in the origin of planets' orbital elements. In fact, 
Terquem \& Papaloizou [2002] provide an important mechanism to explain
the observed planetary systems: massive planets form at the inner part of 
the system and move on a circular orbit due to tidal interaction
with the disc, whereas 
lower mass planets form at the outer part and interact with the
inner ones. Thus, the eccentricities of the 
massive inner planets  increase, which is in accord with the 
observations.
Moreover, the simulations in 
Jiang \& Ip [2001] show that the orbital configuration of  
the planetary system of upsilon Andromedae
might initially be caused by the disc interaction.
Although the discs might be depleted and 
gradually 
become less massive, 
there could be 
belts surviving until now as the Asteroid and Kuiper Belts
in the solar system and these belts would be important for whole dynamical 
histories of planets.



On the other hand,
Thommes, Duncan \& Levison [1999] claimed that Neptune 
was closer to Jupiter and got scattered
outward. 
Yeh \& Jiang [2001] studied the orbital 
migration of scattered planets and completely classified the 
parameter space and the solutions.
They concluded that the eccentricity always increases if the planet which 
moves on a circular orbit initially, is scattered to migrate outward.
Thus, the orbital circularization must be important for the scattered planets
if they are now moving on nearly circular orbits. 

Since belts of planetesimals often 
exist within a planetary system and provide the possible 
mechanism of orbital circularization,
it is important to understand the solutions of dynamical systems with 
planet-belt interaction.
Therefore, Jiang \& Yeh [2003] did some analysis
of the solutions for dynamical systems of 
planet-belt interaction for a belt    
with simple density profile $c/r$, where
$c$ is a constant. However, because the gravitational force from the belt 
involves complicated elliptic integrals and is difficult for the construction
of theorems, they used simplified analytic formulas to estimate this force.
Their main conclusion is that the inner edge of the belt is more stable
than the outer edge of the belt for the planets. 
Their result
is actually consistent with the observational picture of the Asteroid Belt. 

 
In this paper, we complete the work initiated in our earlier paper
Jiang \& Yeh [2003], by numerically evaluating elliptic integrals
to obtain the gravitational force of the
belt without introducing any simplifications. 
Thus, this study is intended to improve 
the understanding about orbital evolution of a given planet-belt system
and hopefully explain why there are less
Kuiper Belt Objects larger than 160 km in diameter beyond 50 AU 
in the outer Solar System.
In Section 2,
we mention our basic governing equations
and in Section 3, we study the locations of the fixed points.
The phase-plane analysis is discussed 
in Section 4 and Section 5 concludes the paper.


\section{The Model}

The belt is an annulus with inner radius $r_1$ and
outer radius $r_2$, where  $r_1$ and $r_2$ are assumed to be 
constants. According to Lizano \& Shu [1989], the belt's surface density  
scales with distance as $r^{-2}$.  We thus assume 
the density profile of the belt to be
$\rho(r)={c}/{r^2}$, where $c$ is a constant, completely
determined by the total mass of the belt. One might notice that there 
are discontinuities at $r=r_1$ and $r=r_2$. This is probably acceptable
because the edges of the belt are rather sharp
as one can see from a plot of the Asteroid Belt in Murray \& Dermott [1999].
 
 
The units of length and time are chosen so as to easily 
compare with our Solar System. 
There are two belts for our Solar System: the Asteroid Belt
in the inner part of the Solar System and the Kuiper Belt
in the outer part. We set $r_1=3$ and $r_2=6$ because (1) when 
the length unit is 1 AU (and the time unit is $1/(2\pi)$  years),
the interval $[r_1, r_2]$ covers the region of the Asteroid Belt
approximately, (2) when 
the length unit is 10 AU (and the time unit is $10^{3/2}/(2\pi)$  years),
the interval $[r_1, r_2]$ covers the region of the Kuiper Belt.


We assume the distance between the central star and the planet to be $r$,
where  $r$ is a function of time. 
The total force $f$ acting on the planet, 
includes the contributions from the central star 
and the belt. The gravitational force from the central star is 
\beq
f_s=-\frac{m}{r^2},\label{eq:fs}
\eeq
where we have set the mass of the central star to be 1 and also 
the gravitational constant $G=1$. 
  

On the other hand, the gravitational force from the belt for the planet is 
complicated and involves
elliptic integrals. We use the equations in Appendix A to get 
the force $f_b$ numerically. 

Because there might be some scattering between the planetesimals in the  
belt and the planet,  the planet should experience the 
frictional force when it goes into the belt region. 
This frictional force 
should be proportional to the surface density of the belt at the location
of the planet, and the velocity of the planet. However, if the planet is doing 
circular motion, the probability of close encounter between the planet and
the planetesimals in the belt is very small and can be neglected here. Thus,
we can assume that the frictional force is proportional to the radial
velocity of the planet  $dr/dt$ only, and ignore the $d\theta/dt$  dependence.

Therefore,  the frictional force  
$\tilde{f}_{\alpha}$ is proportional to the surface density of the 
belt and radial velocity $dr/dt$. Hence, we write down the formula for 
frictional force as
\beq
\tilde{f}_\alpha=-\alpha\rho(r)\frac{dr}{dt}, \label{eq:force1}
\eeq
where $\alpha$ is a frictional parameter and $\rho$ is the density profile of 
belt.

Because the  $d\theta/dt$  component of the planetary velocity 
is ignored in the frictional force,  
the angular momentum $l$ is conserved here, so we have 
\beq
mr^2{d\theta}=l{dt}. \label{eq:angu_mom}
\eeq
This implies that Kepler's second law is still valid here
and, because of this, we can use $\theta$ as our independent variable.
We use $\theta$ to label time $t$ from now on and one can easily
get $t$ from the above equation. 

Since $u=1/r$ and using Equation (\ref{eq:angu_mom}), we have 
\beq 
\frac{dr}{dt}=\frac{dr}{d\theta}\frac{d\theta}{dt}
=\frac{l}{mr^2}\frac{dr}{d\theta}=-\frac{l}{m}\frac{du}{d\theta}.
\label{eq:drdt}
\eeq

Equation (\ref{eq:force1}) can be rewritten as

\beq
\tilde{f}_\alpha
=-\alpha\rho(r)\left[-\frac{l}{m}
\frac{du}{d\theta}\right]
=\frac{\alpha l \rho}{m}\frac{du}{d\theta}.
\eeq

Since the planet feels the  
frictional force only when it enters the belt region, we define  
\beq 
f_\alpha=\beta\tilde{f}_\alpha,
\eeq
where
\begin{displaymath}
\beta=\left\{
\begin{array}{ll}
1 & {\rm if} \quad r_1<r<r_2,  \\
 0& {\rm if}\quad r>r_2 \quad {\rm or}\quad r<r_1.  
\end{array}\right.
\end{displaymath}

In general, the equation of motion for the planet is (Goldstein [1980])
\beq
\frac{d^2u}{d\theta^2}=-u-\frac{mf\left(\frac{1}{u}\right)}{l^2u^2},
\label{eq:1}
\eeq
where $u=1/r$  and  
$m$, $l$ are the mass and the angular momentum of the planet and
$f$ is the total force acting on the planet. 
This equation is only valid when there is no non-radial force.
We use the polar coordinates $(r,\theta)$ to describe the 
location of the planet.

Because $f=f_s+f_{b}+f_\alpha$,  we transform Equation (\ref{eq:1}) 
into  
\beq
\frac{d^2u}{d \theta^2}=-u+\frac{m^2}{l^2}-\frac{m^2f_b}{l^2u^2}
-\frac{\beta\alpha  \rho}{lu^2}\frac{du}{d\theta}.
\label{eq:p_ode1}
\eeq

Therefore, the equation of motion for this system can be written as 

$$ \frac{du}{d\theta}=v\equiv g_1(u,v), $$
\beq
\frac{dv}{d\theta}=-u-\beta k_1\frac{\rho}{u^2}v+k_2-k_2\frac{f_b}{u^2}\equiv 
g_2(u,v), 
\label{eq:gen_p_ode2}
\eeq
where $k_1\equiv {\alpha}/{l}$, $k_2\equiv m^2/l^2$.

\section{Fixed Points}
 
The fixed points $(u,v)$ of Equation (\ref{eq:gen_p_ode2}) satisfy 
the following equations :

$$v=0,$$
\beq
-u-\beta k_1\frac{\rho}{u^2}v+k_2-k_2\frac{f_b}{u^2}=0. \label{eq:fixed1}
\eeq

From Equation (\ref{eq:fixed1}), the fixed points $(u,0)$ satisfy 
\beq
u^3-k_2u^2+k_2f_b=0.
\eeq

The fixed points can be numerically determined
for any given $k_2$ and $f_b$ and it is obvious that 
the location  of fixed points does not depend on the friction.
Since at fixed points $v=du/d\theta=0$, 
the fixed points correspond to circular orbits.
 Particularly, when $f_b=0$, by 
Equation (11), the fixed point satisfies $u=k_2$ and  
this actually corresponds to
the Keplerian orbit. (One can check easily by writing 
$k_2$ as a function of usual angular velocity.) 

The density profile of the belt is 
$\rho(r)={c}/{r^2}$, where $c$ is
determined by the total mass of the belt (Lizano \& Shu [1989]).
Hence, 
the total mass of the belt is 
\beq
M_{b}=\int^{2\pi}_{0}\int^{r_2}_{r_1}\rho(r')r'dr'd\phi =2\pi c
(\ln r_2-\ln r_1). 
\eeq

From Appendix A, we can then calculate 
$f_b$ for any given total mass of the belt $M_b$ as the constant $c$
can be uniquely determined from $M_b$. Moreover, 
since $m$ is always $0.001$ (about $M_J $) in this paper, 
$k_2$ is directly related to the angular momentum of the planet. 
Therefore, we use $M_b$ and $k_2$ 
as the two main parameters to determine the number
and locations of fixed points in this system.


Figure 1 is the bifurcation diagram of fixed points and 
these curves show the locations of fixed points $u$ as a function of $k_2$ for
a given $M_b$.  From this plot, we know that when the total mass
of the belt is very small, $M_b=0.01$, the curve on the 
$k_2-u$ plane is almost 
a straight line with slope 1. That is, there is one and only one fixed
point $u \sim k_2$. 
This can be easily
verified by Equation (11) with small $f_b$, and it corresponds to
the Keplerian orbits. When the mass of the belt is not negligible,
the curves form two needle-like structures pointing along
both, the inner edge of the belt, where $u=1/r_1=1/3$, and the 
outer edge of the belt, where $u=1/r_2=1/6$, and there would be more 
than one fixed point for particular values of $k_2$. In addition to the 
one corresponding to the Keplerian orbit, other fixed points are 
located around inner and outer edges of the belt. Thus, 
the belt could bring the planets to non-Keplerian circular orbits by the
gravitational and frictional forces. 

When $M_b=0.1$, the  needle-like structures
imply  the existence of two 
fixed points for $0.5 < k_2 < 0.7$. When $M_b=0.5$, the two needle-like 
structures  
become even bigger and there are three fixed points for $0.5 < k_2 < 1$
, two fixed points around $k_2=0.06$ and one fixed point for most other 
values of $k_2$.
When $M_b=1$, the structure grows even larger and becomes completely different
from a straight line.  There are three fixed points for larger $k_2$ 
($k_2 > 0.6$) and 
one fixed point only for both intermediate $k_2$ ($0.1 < k_2 < 0.5$) 
and tiny $k_2$ ($k_2 < 0.02$). Since the needle-like structure around the 
outer edge is tiny, the probability that the planet moves stably around the 
outer edge is much smaller than near the inner edge.

\section{Phase-Plane Analysis}
Following the linearization analysis, the eigenvalues 
$\lambda$ corresponding to the fixed points $(u_\ast,0)$ satisfy the following 
equation:
\beq
\left(\frac{\pa g_1}{\pa u}\biggm|_{(u_\ast,0)} -\lambda\right)
\left(\frac{\pa g_2}{\pa v}\biggm|_{(u_\ast,0)} -\lambda\right)-
\left(\frac{\pa g_1}{\pa v}\biggm|_{(u_\ast,0)}\right)
\left(\frac{\pa g_2}{\pa u}\biggm|_{(u_\ast,0)}\right)=0,
\eeq
where $g_1(u,v)=v$ and $g_2(u,v)=-u-\beta{(k_1\rho v)}/{u^2}+k_2-{(k_2f_b)}
/{u^2}$. Hence, the above equation becomes
 
$$\lambda^2+\frac{\beta k_1\rho}{u_\ast^2}\lambda+\left(1
-\frac{2\beta k_1\rho v}{u^3}
\biggm|_{(u_\ast,0)}+k_2\frac{\pa A}{\pa u}\biggm|_{(u_\ast,0)}\right)=0,$$
\beq
 {\rm where}\,\,  A=\frac{f_b}{u^2}.\label{eq:a1}
\eeq

When the planet is out of the belt region, that is $\beta=0$, we have 
$$\lambda=\pm\sqrt{-1-k_2\frac{\pa A}{\pa u}\biggm|_{(u_\ast,0)}},$$
and there are two possible cases:

Case I: If $-1-k_2\frac{\pa A}{\pa u}\biggm|_{(u_\ast,0)}>0$, this fixed point
            is a saddle point. 

Case II: If $-1-k_2\frac{\pa A}{\pa u}\biggm|_{(u_\ast,0)}<0$, this fixed 
         point is a center point.

 When the planet stays in the belt, that is $\beta=1$, we have 
$$\lambda=\frac{k_1\rho}{2u_\ast^2}\left\{-1\pm\sqrt
{1-4\left(1+k_2\frac{\pa A}{\pa u}\biggm|_{(u_\ast,0)}\right)
\left(\frac{u_{\ast}^2}{k_1\rho}\right)^2}\right\}.
$$
We define $\Delta={1-4\left(1+k_2\frac{\pa A}{\pa u}\biggm|_{(u_\ast,0)}\right)
\left(\frac{u_{\ast}^2}{k_1\rho}\right)^2}$, then we have 
three possible cases:

Case 1: When $\Delta <0$, two eigenvalues $\lambda_{1,2}=-a\pm b{\rm i}$ 
for  $a,b>0$, so the fixed point is a stable spiral point. 

Case 2: When $0<\Delta <1$, both eigenvalues are negative, so the fixed 
point is a stable node. 

Case 3: When $\Delta >1$, one eigenvalue is negative and the other is 
positive, so the fixed point is a saddle point. 


We understand that when the real parts of the eigenvalues of a 
fixed point equal to zero in the first-order linearization analysis, 
the properties of this fixed point should be determined by the higher
order analysis in general. However, to simplify our language, we still 
use the term  ``center point'' for any fixed point 
having an eigenvalue 
with zero real part.
This is a good choice because our numerical results
indicate that these points are in fact the center points.

We now derive the mathematical formula for $\Delta$. At first, from 
Equation (\ref{eq:a1}), we calculate (please see details in Appendix B)
\beqn
\frac{\pa A}{\pa u}\biggm|_{(u_\ast,0)}&=&
\frac{1}{u_\ast^2}\frac{\pa f_b}{\pa u}\biggm|_{(u_\ast,0)}-\frac{2f_b}
{u_\ast^3}
=-\frac{1}{u_\ast^4}\frac{\pa f_b}{\pa r}\biggm|_{(u_\ast,0)} 
-\frac{2f_b}{(u_\ast)^3}, \non \\ 
&=&-\frac{2}{u_\ast}\int_{r'} r'\rho(r')\left\{\frac{1+r'^2u_\ast^2}
{(1-r'u_\ast)^2(1+r'u_\ast)} E-\frac{1}{1+r'u_\ast}F\right\} dr',
\eeqn
where  $E$ is an elliptic integral of the second kind and  $F$ is 
an elliptic integral of
the first 
kind. Hence, we have
\beqn
\Delta&=&1-4\left(\frac{u_\ast^2}{k_1\rho}\right)^2
\left(1+k_2\frac{\pa A}{\pa u}\biggm|_{(u^\ast,0)}\right) \non \\
&=& 1-4\left(\frac{u_\ast^2}{k_1\rho}\right)^2
\left\{1-\frac{2k_2}{u_\ast}\int_{r'} r'\rho(r')\left[\frac{1+r'^2u_\ast^2}
{(1-r'u_\ast)^2(1+r'u_\ast)} E-\frac{1}{1+r'u_\ast}F\right] dr'\right\} 
\non \\
&=&1-4\left(\frac{u_\ast^2}{k_1\rho}\right)^2+\frac{8u_\ast^3k_2}{k_1^2\rho^2}
\int_{r'} r'\rho(r')\left[\frac{1+r'^2u_\ast^2}
{(1-r'u_\ast)^2(1+r'u_\ast)} E-\frac{1}{1+r'u_\ast}F\right] dr'.
\eeqn

If we calculate the value of $\Delta$ numerically, we can then understand the 
properties of this fixed point. For example, in Figure 2-6,
we choose four sets of 
($M_b$, $k_2$) values
and  make the phase-plane plots, orbits  etc. 
to demonstrate the orbital evolution, and the stability of the system
to provide a general picture for the readers. In the results below, 
the frictional parameter 
$\alpha$ were chosen to be small numbers but large enough to 
make the solution curve 
not too tight to be studied.
The value of the frictional parameter 
$\alpha$ is related to the 
viscosity of the belt and is too complicated to be well-determined 
observationally. 
 The values of $\alpha$ in the models would affect 
the time scales of orbital evolution but would not change the general 
understanding of the dynamical properties of the system.
For the parameters we choose, some results show that the orbits
become nearly circular in an order of $10^3$ years if the unit of length   
is 10 AU. This is rather short and perhaps only possible during the 
period in which the planetary system is just formed. Nevertheless,
one can easily choose smaller values of $\alpha$ to get results
of orbital evolution with longer time scales when the results are used
to describe more recent systems.

Figure 2 is the case when we set $M_b=0.5$ and $k_2=0.6$. We assume 
the frictional parameter $\alpha=0.001$ for this case.  
We know from Figure 1,  that there must be three fixed points and Figure 2(a) 
shows that the one around $u=0.6$  is a center point and the other two near the
inner edge of the belt are spiral and saddle points, respectively. 
The blue dotted curve around 
the center point in Figure 2(a) is then plotted on the $x-y$ plane as 
the blue dotted orbit in Figure 2(b). It is  
 a precessing elliptical orbit with small eccentricity.  
The red solid curve approaching the spiral 
point in Figure 2(a) is then plotted on the $x-y$ plane as 
the red solid orbit in Figure 2(b). This orbit keeps changing the semimajor 
axis and eccentricity and finally becomes a circular orbit with radius
$r=r_1$, i.e., close to the inner edge of the belt.
Figures 2(c)-(d) show $u$ and $v$ as functions of $\theta$ for both
above orbits. Because $u$ becomes a constant and $v$ approaches to zero 
when $\theta > 30$ for the 
red solid curves in Figures 2(c)-(d), these two figures
reconfirm that the red solid curve in Figure 2(b) finally becomes a circular 
orbit. While, since both $u$ and $v$ keep oscillating for the blue dotted 
curves in Figures 2(c)-(d),
the blue dotted curve in Figure 2(b) is then a precessing elliptical orbit.

Figure 3 is the case when we set $M_b=0.1$ and $k_2=0.3$. 
We assume the frictional parameter $\alpha=0.01$ for this case.  
From Figure 1, we know that there is only one fixed point and it is 
close to the inner edge of the belt, where $u=1/r_1=1/3$. 
Figure 3(a) shows that this fixed point is a spiral point. The solution curve 
on the $u-v$ phase-plane is then plotted as an orbit on the $x-y$ plane
in Figure 3(b).   We can see that the planet moves across the belt
initially and gradually settles down on a circular orbit close to the 
inner edge of the belt.
Figures 3(c)-(d) show $u$ and $v$ as functions of $\theta$ for this orbit.

Figure 4 is the case when we set $M_b=0.5$ and $k_2=0.06$. We assume 
the frictional parameter $\alpha=0.1$ for this case.  
We know, from Figure 1,  that there are two fixed points. Figure 4(a) 
shows that the one around $u=0.1$  is a center point and another near 
the outer edge of the belt is a spiral point. The blue dotted curve around 
the center point in Figure 4(a) is then plotted on the $x-y$ plane as 
the blue dotted orbit in Figure 4(b). It is obviously 
 a precessing elliptical orbit.  
The red solid curve approaching the spiral 
point in Figure 4(a) is then plotted on the $x-y$ plane as 
the red solid orbit in Figure 4(b). This orbit keeps changing the semimajor 
axis and eccentricity and finally becomes a circular orbit with radius
$r=r_2$, i.e., close to the outer edge of the belt.
Figures 4(c)-(d) show $u$ and $v$ as functions of $\theta$ for both
above orbits.


Figures 5-6 are for the case when we set $M_b=0.1$ and $k_2=0.1$. We assume 
the frictional parameter $\alpha=0.1$ for this case.  
We know from Figure 1,  that there is only one fixed point and Figure 5(a) 
shows that this fixed point is a center point.
We found that   
there is a ``limit cycle'' that all solution curves out of this cycle
are approaching this cycle. However, all the solution curves within this cycle
do not, they still behave as normal curves around a center point.
Therefore, this kind of ``limit cycle'' is called a ``semi-attractor''.

To understand how these outer solution curves approach this
 ``limit cycle'', we simply pick up 
the blue dotted curve's representative points which are on the right hand side 
of the fixed point and cross $v=0$ line, i.e. the points satisfying 
$u > 0.12$ and $v=0$
in Figure 5(a). Figure 5(b) is the plot of the $u$ values of these 
points, $\bar u$, as a function of $\theta$ when we set initial $\theta=0$. We
can see that $\bar u$ approaches to ${\bar u}=1/r_2=1/6$ 
and the approaching speed 
decays very quickly.
In Figure 5(b), we see that the curve almost becomes
a straight line, i.e. $d {\bar u}/d {\theta}=0$  when $\theta > 100$.

The blue dotted curves in Figure 5(c)-(d) show $u$ and $v$ as functions of 
$\theta$ and the curve in Figure 6(a) shows the orbit on the $x-y$ plane
for the above solution curve which approaches the ``limit cycle''.
Figure 6(b) is the closer view of Figure 6(a). 

Moreover,
the red solid curves in Figure 5(c)-(d) show the $u$ and $v$ as function of 
$\theta$ and the curve in Figure 6(c) shows the orbit on the $x-y$ plane
for the red solid solution curve around the center point in Figure 5(a).

Figure 6(d) is another orbit for a given solution curve approaching the
``limit cycle''.  The orbit is getting closer to 
 a precessing ellipse when the solution curve approaches the
``limit cycle'' as we can see in both Figure 6(b) and Figure 6(d). 


\section{Conclusions}

We have studied the orbital evolution and also stability of planetary systems 
with interaction from belts by the standard phase-plane analysis. 
We regard the mass of the belt $M_b$ and $k_2$ ($= m^2/l^2$) as the two 
main parameters to determine the location of fixed points and also
classify the outcome of orbital evolution. Please note that the fixed points
on the phase space corresponds to the circular orbits on the orbital plane.

From the results in Figure 1, we know that when the belt's mass is small, 
the bifurcation diagram is almost a straight line, i.e. Keplerian orbits.  
When the belt's mass is larger, there could 
be more than one fixed point, one still corresponding 
to the Keplerian orbit and
the new fixed points always close to the inner edge  $u=1/r_1$ 
or outer edge $u=1/r_2$ of the belt.

 However, since the needle-like 
structure
around the outer edge in Figure 1 
is very tiny, the probability that the planet 
moves stably around the outer edge is much smaller than near the inner edge.
This conclusion is consistent with the principle result in Jiang \& Yeh [2003].
 
There is one interesting case in which 
we found a limit cycle of semi-attractor type:
the solution curves lying outside of the cycle
approach this cycle but the ones within this cycle do not.
In this case, the orbit is getting similar to 
a precessing ellipse when the solution curve approaches the
limit cycle. 


What could we learn for the Solar System from these theoretical results ? 




It is  known that there is a belt, the Kuiper Belt, in the outer Solar System 
after the first Kuiper Belt Object (KBO)
was detected (Jewitt \& Luu [1993]).
Many more KBOs have been detected since then and now 
there are more than 500 discovered KBOs.
Allen, Bernstein \& Malhotra [2001]
did a survey and found that they could not find any KBOs
larger than 160 km in diameter beyond 50 AU in the outer solar system.
They suggested several possible explanations for this observational result. 

If we apply our model to this problem,  the point mass which represents
the planet in our model, now represents a KBO. Our results imply that 
the big KBOs would have greater probability of approaching the inner edge
of the Kuiper Belt and the small KBOs would not.
This is because the value of $k_2$ would be larger
when $m$ is large for a given initial angular momentum, and thus it corresponds
to the right branch of curves in Figure 1. If $m$ is much smaller and 
 $k_2<0.1$, it corresponds to the left branch of curves in Figure 1
and thus the small KBOs might not approach the inner edge of the Kuiper Belt.


Therefore, 
our result is consistent with one of the possible reasons: 
larger KBOs might have been 
displaced to their present orbits by a large-scale rearrangement of orbit 
(Allen, Bernstein \& Malhotra [2001]). 
In fact, our results  provide a natural mechanism 
to do this orbit rearrangement:
larger KBOs might have been moving towards the inner edge of the belt 
due to the friction from the belt.  

\section*{Acknowledgment}
We are grateful to both the referees for their suggestions.
This work is supported in part 
by the National Science Council, Taiwan, under Grants NSC 90-2112-M-008-052.


\clearpage

\appendix
\section{Gravitational Force from the Belt $f_b$}

Let $F(\xi)$ be an elliptic integral of the first kind  and assume
\beq
\xi=\frac{2\sqrt{rr'}}{r+r'}.\label{eq:xi}
\eeq
By  $\sin^2 A=\frac{1}{2}(1-\cos 2A)$, we have 
\beq
F(\xi)=\int^{\pi/2}_0\frac{1}{\sqrt{1-\xi^2\sin^2\phi'}}d\phi'
=\frac{(r+r')}{2} \int^{\pi}_{0}\frac{1}{\sqrt{r^2+r'^2-2rr'\cos (\phi)}}
d\phi.  \label{eq:f_1}
\eeq 
Therefore, from Equation (\ref{eq:f_1}), we have
\beq
\int^{2\pi}_{0}\frac{1}{\sqrt{r^2+r'^2-2rr'\cos (\phi)}}d\phi 
=2\int^{\pi}_{0}\frac{1}{\sqrt{r^2+r'^2-2rr'\cos (\phi)}}d\phi 
=\frac{4F(\xi)}{r+r'}.\label{eq:f_2}
\eeq

The gravitational potential is 
\beq
V(r)=-G\int^{r_2}_{r_1}\!\int^{2\pi}_{0}
\frac{\rho(r')r'}{\sqrt{r^2+r'^2-2rr'\cos (\phi')}}dr'\!d\phi'
=-4G\int^{r_2}_{r_1} \frac{F(\xi)\rho(r')r'}{r+r'}dr', 
\eeq
where Equation (\ref{eq:f_2}) has been used.

Now we differentiate this potential with respect to $r$ to get
 the gravitational force, so 
\beq
f_b=-\frac{\pa V}{\pa r}=4G\int^{r_2}_{r_1} \frac{\frac{dF(\xi)}{d\xi}
\frac{\pa \xi}{\pa r}\rho'(r')r'}{r+r'}dr'+4G\int^{r_2}_{r_1} 
F(\xi)\rho(r')r'\frac{\pa}{\pa r}\left(\frac{1}{r+r'}\right)dr'.
\label{eq:dvdr}
\eeq
Let $E(\xi)$ be an elliptic integral of the second kind and 
$\xi'=\sqrt{1-\xi^2}$.
Since $F(\xi)$ is  an elliptic integral  of the first kind, from 
Byrd \& Friedman [1971],
we have
\beq
\frac{d F(\xi)}{d\xi}=\frac{E-\xi'^2F(\xi)}{\xi\xi'^2}=
\frac{E}{\xi(1-\xi^2)}-\frac{F}{\xi}.\label{eq:dfdxi}
\eeq

By Equation (\ref{eq:xi}) and (\ref{eq:dfdxi}), we calculate
\beq
\frac{dF(\xi)}{d\xi}\frac{\pa \xi}{\pa r}=\left[
\frac{E}{\xi(1-\xi^2)}-\frac{F}{\xi}\right]\sqrt{\frac{r'}{r}}\frac{r'-r}
{(r+r')^2} 
= -\frac{E(r+r')}{2r(r-r')}+\frac{F(r-r')}{2r(r+r')}. \label{eq:dfxi_dxidr}
\eeq 

We substitute Equation (\ref{eq:dfxi_dxidr}) into 
Equation (\ref{eq:dvdr}), so we have
\beq
f_b=-\frac{\pa V}{\pa r}
= -2G\int^{r_2}_{r_1}\frac{\rho(r')r'}{r}\left[\frac{1}{r-r'}E+\frac{1}{r+r'} F
\right]dr'.\label{eq:f_b}
\eeq
If the planet is on the belt region, 
there is a singularity for $f_b$ and 
we introduce a small  $\epsilon>0$ to avoid it. 
That is, we numerically 
calculate the force from the belt by the following formula:
\beq
f_b= -2G\int^{r-\epsilon}_{r_1}\frac{\rho(r')r'}{r}\left[\frac{1}{r-r'}E
+\frac{1}{r+r'} F\right]dr'-2G\int^{r_2}_{r+\epsilon}
\frac{\rho(r')r'}{r}\left[\frac{1}{r-r'}E+\frac{1}{r+r'} F
\right]dr'.
\eeq

Similar treatment can be done for the integrals in Appendix B.

\section{Derivatives of $f_b$}

From Equation (\ref{eq:f_b}), we calculate 

\beqn
\frac{\pa f_b(r)}{\pa r}&=&\frac{2G}{r^2}\int^{r_2}_{r_1}
r'\rho(r')\left[\frac{1}{r-r'}E+\frac{1}{r+r'} F\right]dr' \non \\
& &-\frac{2G}{r}
\int^{r_2}_{r_1}r'\rho(r')\left[\frac{\frac{dE}{d\xi}\frac{d\xi}{dr}}{r-r'}
+E\frac{d}{dr}\left(\frac{1}{r-r'}\right)+\frac{\frac{dF}{d\xi}
\frac{d\xi}{dr}}{r+r'}+F\frac{d}{dr}\left(\frac{1}{r+r'}\right)\right]dr'.
\label{eq:fb_dr}
\eeqn

From Byrd \& Friedman (1971),
$$\frac{dE}{d\xi}=\frac{1}{\xi}(E-F)=\frac{r+r'}{2\sqrt{rr'}}(E-F).$$
Thus, 
\beq
\frac{dE}{d\xi}\frac{d\xi}{dr}=\frac{r'-r}{2r(r+r')}(E-F).
\label{eq:dexi_dxidr}
\eeq

We substitute Equation (\ref{eq:dfxi_dxidr}) 
and Equation (\ref{eq:dexi_dxidr}) 
into Equation (\ref{eq:fb_dr}), so we have

\beq
\frac{\pa f_b(r)}{\pa r}
=\frac{2G}{r^2}\int^{r_2}_{r_1}r'\rho(r')\left\{
\frac{3r^2-r'^2}{(r-r')^2(r+r')} E+\frac{1}{r+r'}F\right\} dr'.
\eeq

Hence, from Equation (\ref{eq:f_b}), we have  
\beq  
\frac{1}{u^4}\frac{\pa f}{\pa r}+\frac{2f_b}{u^3}
=\frac{2G}{u}\int^{r_2}_{r_1}r'\rho(r')\left\{
\frac{1+r'^2u^2}{(1-r'u)^2(1+r'u)} E-\frac{1}{1+r'u}F\right\} dr'.
\eeq

\clearpage

\begin{figure}[tbhp]
\epsfysize 5.5in \epsffile{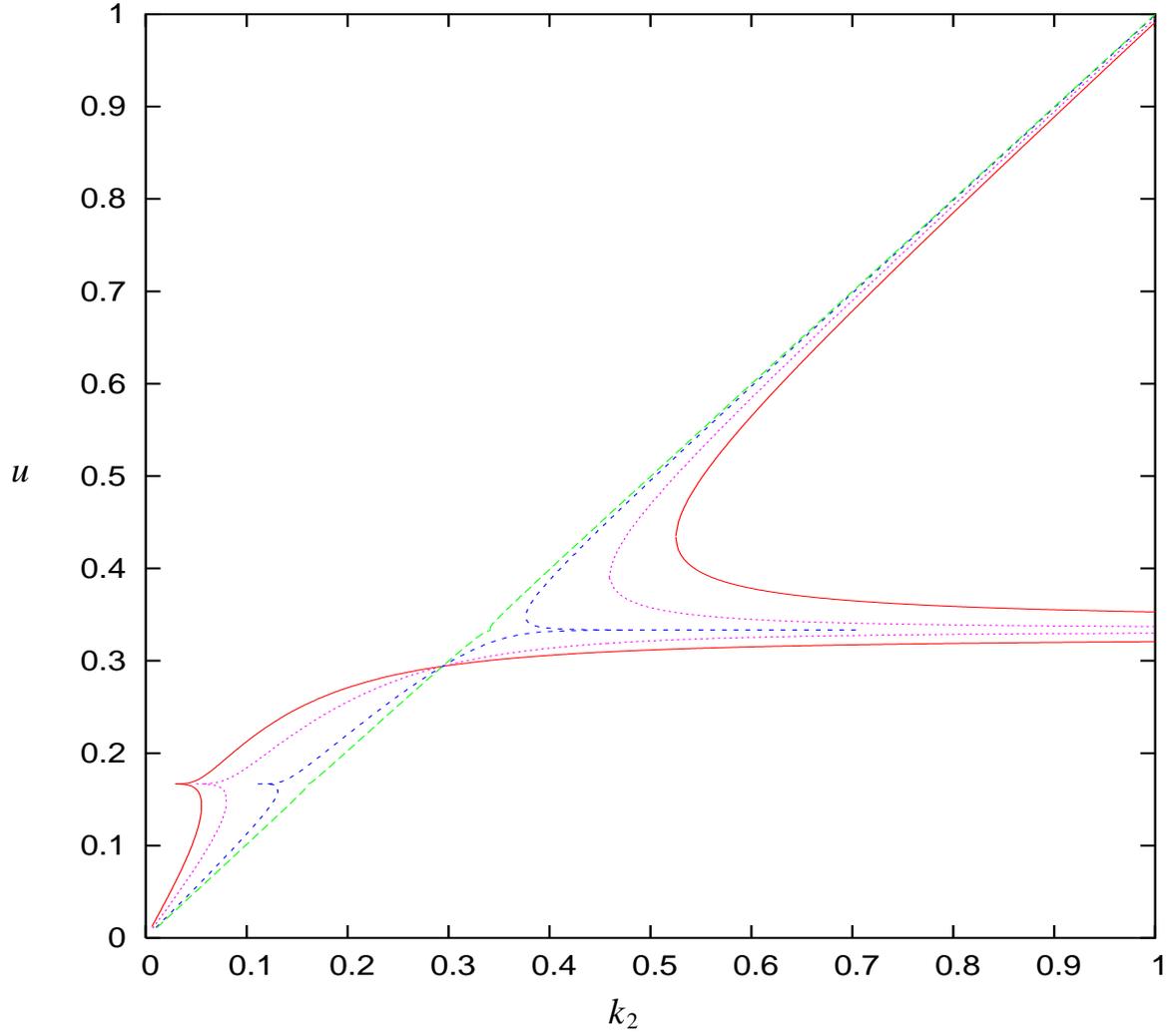}
\caption{The
bifurcation diagram of fixed points on the $k_2-u$ plane with $r_1=3$, $r_2=6$,
where the green dashed curves are for $M_b=0.01$, 
the blue dotted curves are for $M_b=0.1$,
the purple dotted curves are for $M_b=0.5$ and 
the red solid curves are for $M_b=1$. 
}
\end{figure}

\clearpage

\begin{figure}[tbhp]
\epsfysize 7.0in \epsffile{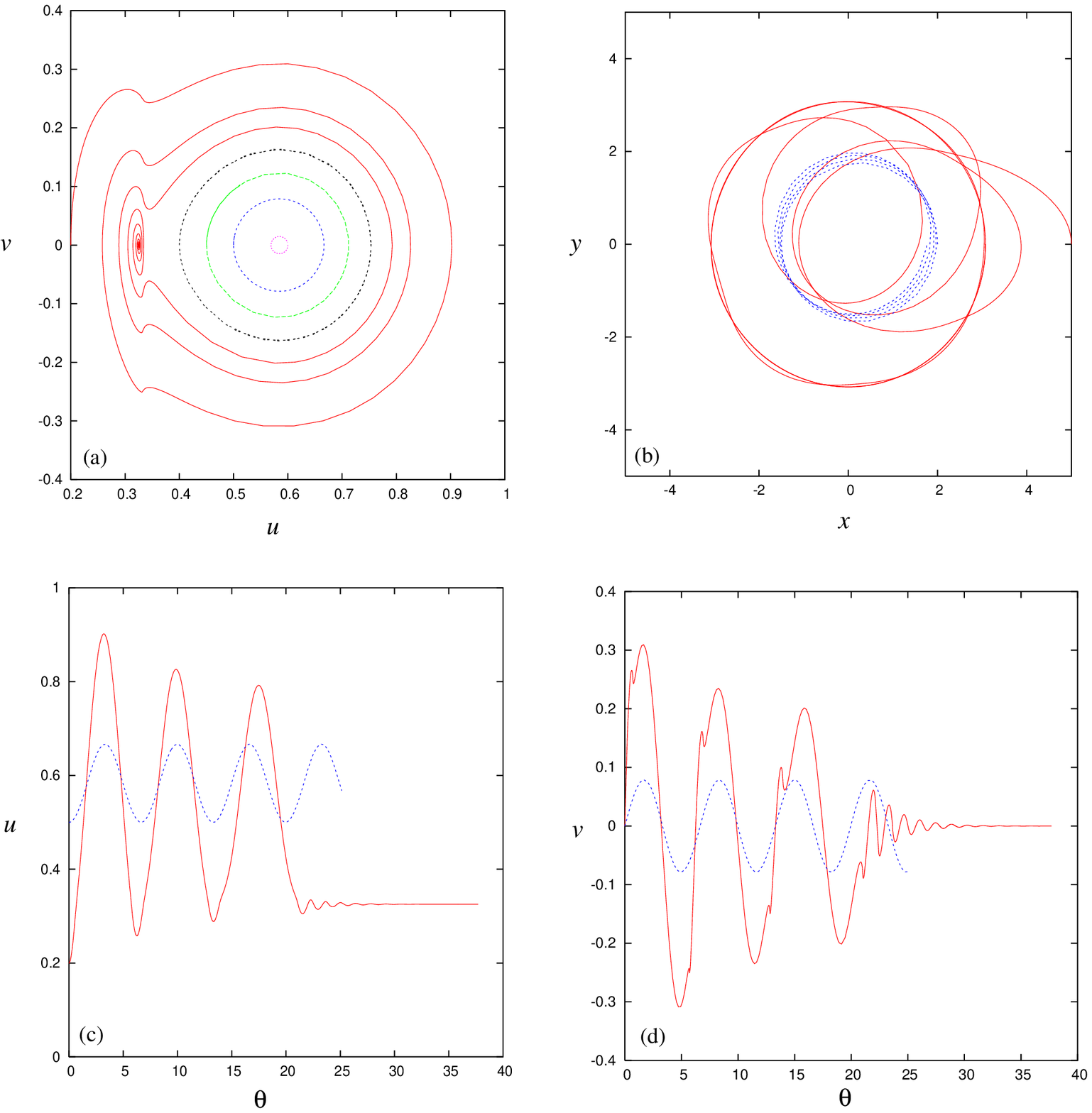}
\caption{The result for $M_b=0.5$ and $k_2=0.6$ with 
$\alpha=0.001$, $r_1=3$, $r_2=6$.
(a) Solution curves on the $u-v$ phase-plane. 
(b) Orbits on the $x-y$ plane. 
(c) The plots of $u$ as a function of $\theta$. 
(d) The plots of $v$ as a function of $\theta$.
The red solid (blue dotted) 
curves in all panels correspond to the same solution.
}
\end{figure}

\clearpage

\begin{figure}[tbhp]
\epsfysize 7.0in \epsffile{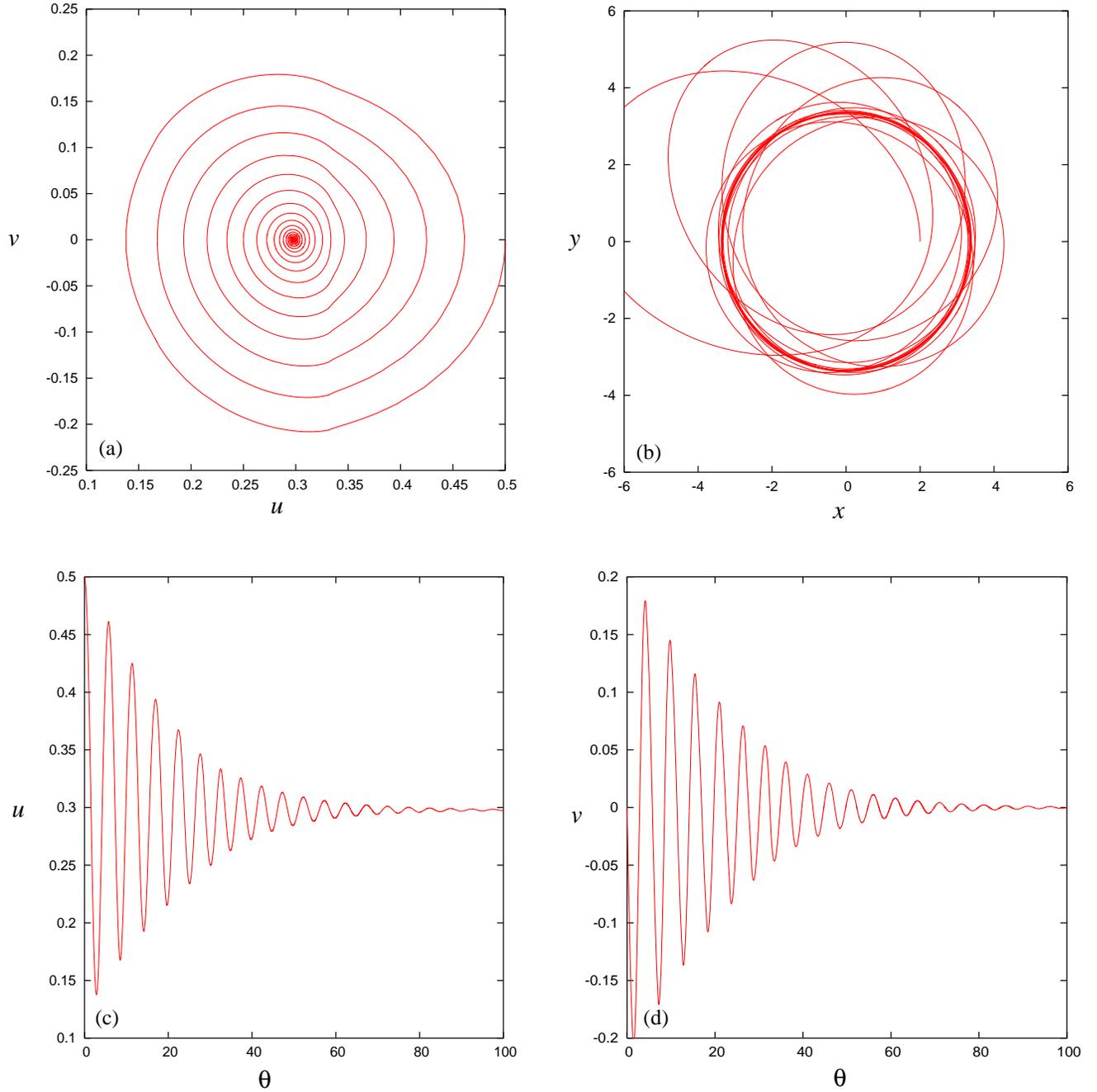}
\caption{The result for $M_b=0.1$ and $k_2=0.3$ with 
$\alpha=0.01$, $r_1=3$, $r_2=6$.
(a) A solution curve on the $u-v$ phase-plane. 
(b) An orbit on the $x-y$ plane. 
(c) The plot of $u$ as a function of $\theta$. 
(d) The plot of $v$ as a function of $\theta$.
The curves in all panels correspond to the same solution.
}
\end{figure}

\clearpage

\begin{figure}[tbhp]
\epsfysize 7.0in \epsffile{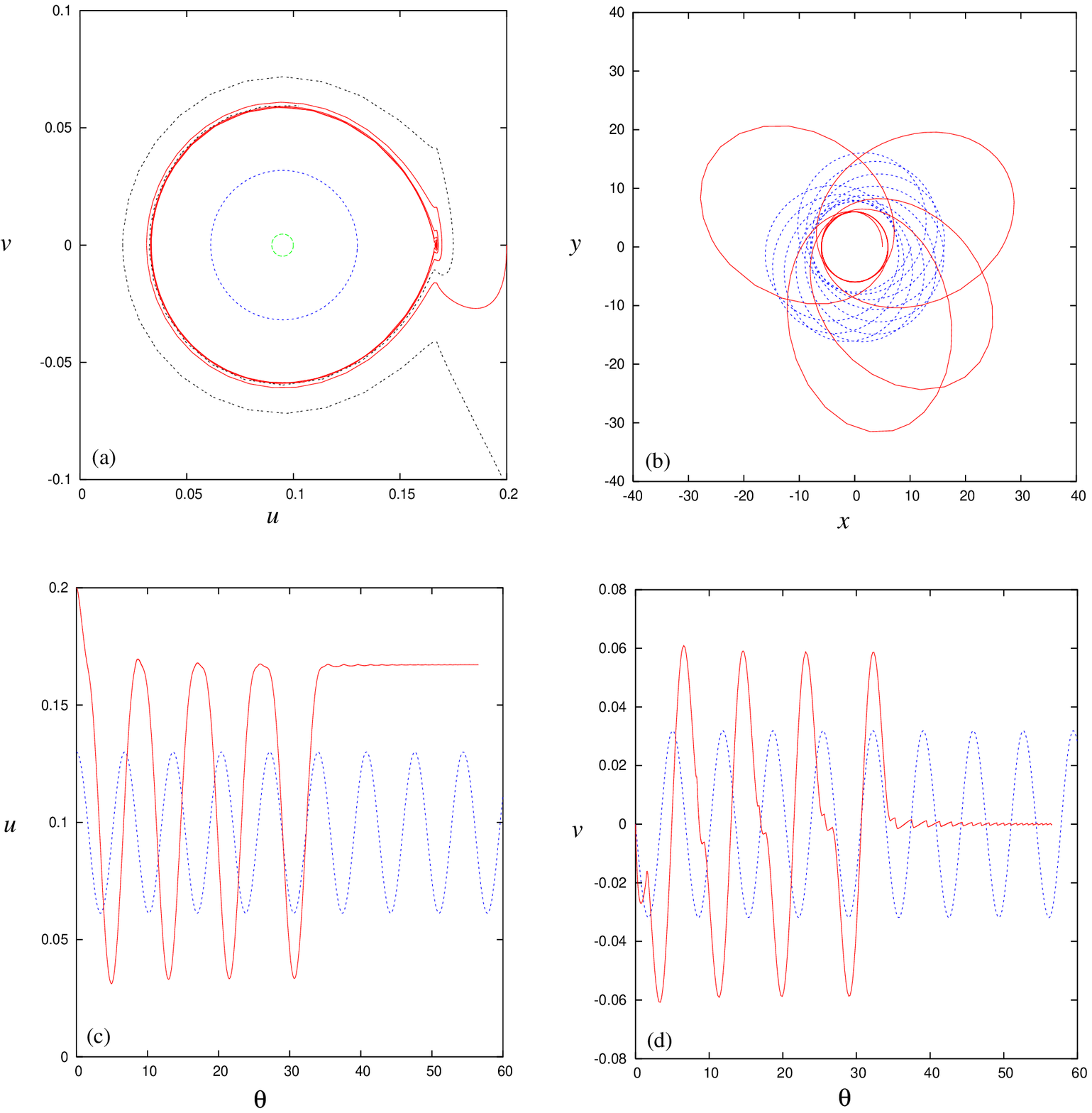}
\caption{The result for $M_b=0.5$ and $k_2=0.06$ with 
$\alpha=0.1$, $r_1=3$, $r_2=6$.
(a) Solution curves on the $u-v$ phase-plane. 
(b) Orbits on $x-y$ plane. 
(c) The plots of $u$ as a function of $\theta$. 
(d) The plots of $v$ as a function of $\theta$.
The red solid (blue dotted) 
curves in all panels correspond to the same solution.
}
\end{figure}

\clearpage

\begin{figure}[tbhp]
\epsfysize 7.0in \epsffile{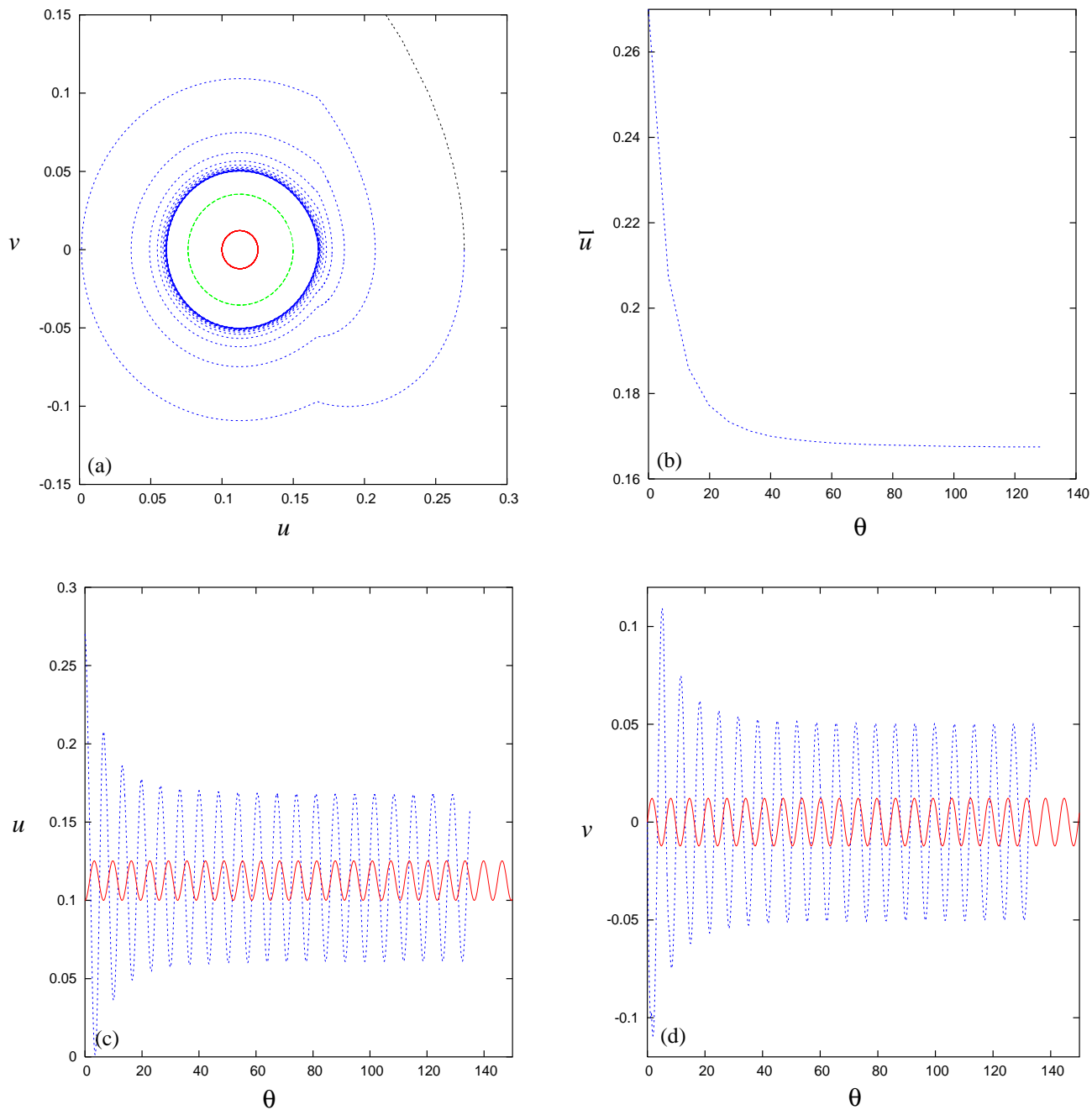}
\caption{
The result for $M_b=0.1$ and $k_2=0.1$ with 
$\alpha=0.1$, $r_1=3$, $r_2=6$.
(a) Solution curves on the $u-v$ phase-plane. 
(b) $\bar u$ as a function of $\theta$, where $\bar u$ 
are the values of $u$ of the blue dotted curve in (a) and satisfied
$u > 0.12$ and $v=0$.
(c) The plots of $u$ as a function of $\theta$. 
(d) The plots of $v$ as a function of $\theta$.
The red solid (blue dotted) 
curves in all panels correspond to the same solution. 
}
\end{figure}

\clearpage

\begin{figure}[tbhp]
\epsfysize 7.0in \epsffile{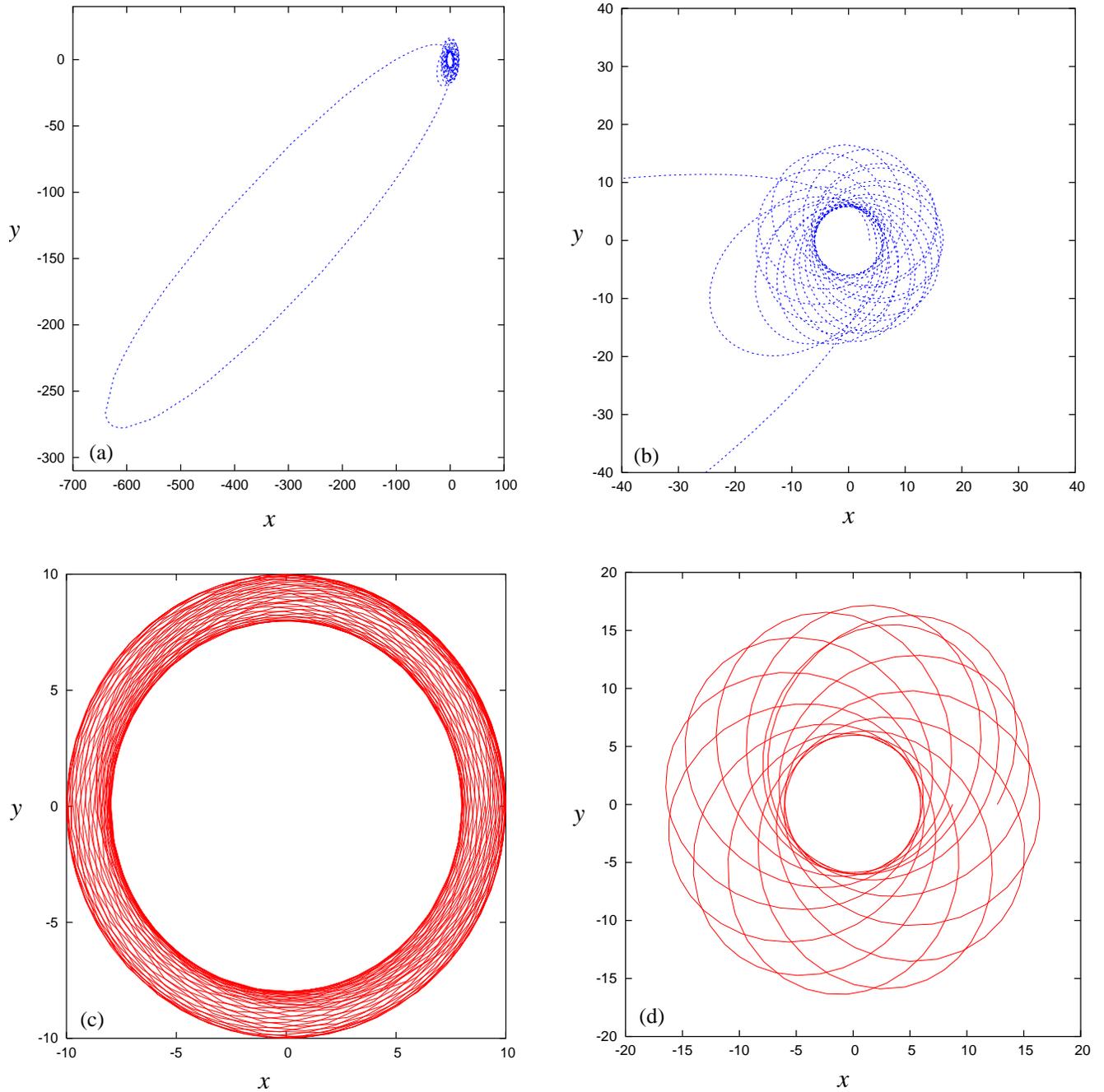}
\caption{The result for $M_b=0.1$ and $k_2=0.1$ with 
$\alpha=0.1$, $r_1=3$, $r_2=6$.
(a) The orbit on the $x-y$ plane for the blue dotted curve in Figure 5(a).
(b) The closer view of (a).
(c) The orbit on the $x-y$ plane for the red solid curve in Figure 5(a).
(d) Another orbit for arbitrary given solution curve approaching 
    the limit cycle. 
}
\end{figure}

\end{document}